\documentclass[12pt]{iopart}
\usepackage{times}
\usepackage{graphicx}

\begin{document}
\title{Thermalization of Squeezed States}
\author{Allan I. Solomon}
\address{Physics and Astronomy Department, The Open University,
Milton Keynes MK7 6AA, U.K. \\{\small and}  \\
LPTMC, University of Paris VI, Paris, France \\
email: a.i.solomon@open.ac.uk}
\begin{abstract}
Starting with a thermal squeezed state defined as a conventional thermal state based on an appropriate hamiltonian, we show how an important physical property, the signal-to-noise ratio, is degraded, and propose a simple model of thermalization (Kraus thermalization).
\end{abstract}

\section{Introduction}
We first of all propose a simple definition of a thermal squeezed state. Previous definitions of thermal coherent states and squeezed states\cite{bar85,man89} have used the thermofield approach of Takahashi and Umezawa\cite{tak75}. An early paper by Vourdas\cite{vour86} discusses the superposition of squeezed states with thermal photons; a more recent paper\cite{bis94} considers only the case of thermal coherent states. An early study of canonically transformed thermal states similar to the approach of the present work was made in \cite{dodo}, and also in \cite{vour87,muss95}. Two further papers adopt essentially the same definition as the present note s\cite{mar1,mar2} using a slightly different approach. The method that we take here is the straightforward one of defining the hamiltonian whose eigenstates are squeezed states (sometimes known as squeezed-coherent states to distinguish them from squeezed vacuum states). The complete set of eigenstates are in fact squeezed displaced number states. These are the states $SD|n\rangle$, where $D$ is the displacement operator, and $S$ the squeezing operator.  The thermal squeezed state we define is simply a  mixed state given as a thermal sum of these states. The simplicity of this approach in, for example, calculating thermal averages is due to the group-theoretical structure of these states, which essentially enables such calculations to be reduced to free hamiltonian averages. This structure is a consequence of the definition of the squeezed-coherent states as transformed vacua under the automorphism group of the canonical commutation relations, which we shall describe in the next section.

To exemplify how the physical properties are degraded by the thermalization process, we calculate the important signal-to-quantum noise ratio for these thermal squeezed states.

Finally, we use the Kraus formalism to propose a model for the thermalization.
\section{Canonical Transformations and Quantum Optics states}
The canonical commutation relations for the boson (photon) creation and annihilation operators $a^{\dagger}$ and $a$ are $[a,a^{\dagger}]=I.$ We shall consider canonical transformations of the form
\begin{equation}
a\rightarrow A = UaU^{\dagger}
\;\;\; \;
\; \;
a^{\dagger}\rightarrow A^{\dagger} = Ua^{\dagger}U^{\dagger}.
\end{equation}
The approach of this note  is to consider the relevant quantum optics states as vacuum states for the transformed operator $A$. The {\em vacuum state} is defined as the normalized state satisfying $ a|0\rangle =0. $ We note that the vacuum state corresponding to $A$, defined by $ A|0'\rangle=0 $ is given by $ |0'\rangle = U|0\rangle. $ In the spirit of our approach, the ordinary vacuum state corresponds to the trivial unitary transformation $U=I$,
\begin{equation}
a\rightarrow A = IaI^{\dagger}
\;\;\;
\;\;\;
a^{\dagger}\rightarrow A^{\dagger} = Ia^{\dagger}I^{\dagger}
\end{equation}
with $ |0'\rangle = I|0\rangle. $
 Less trivially, we shall now consider some important states
which may be considered in the same light.
\begin{itemize}
\item {\bf Coherent States} These states correspond to the (next) most simple canonical transformation, that corresponding to
$$
a\rightarrow A=a-\alpha I
\; \; \; \; \; \;
a^{\dagger}\rightarrow A^{\dagger}=a^{\dagger}-\alpha^{*}I
$$
where $I$ is the unit operator. This canonical transformation is generated by the unitary displacement operator
$$ D(\alpha) =  \exp(\alpha a^{\dagger}-\alpha ^{*} a)$$ which operates on $a$ by
 \begin{equation}
a\rightarrow A = 
D(\alpha)a D^{\dagger}(\alpha) = a-\alpha I.
\label{disp}
\end{equation}
The vacuum state of $A$ is the coherent state
$$\vert 0'
\rangle =
\vert \alpha \rangle = D(\alpha) \vert 0 \rangle$$ and the vacuum
equation for $A$ is $A|0'\rangle=0$ which is equivalent to $ a|\alpha\rangle=\alpha|\alpha\rangle. $ This is the usual coherent state\cite{Glauber}
\begin{equation}
|\alpha\rangle ={\mathcal{N}}(\alpha)
\sum_{n=0}^\infty
     \frac{{\alpha}^n}{\sqrt{n!}} |n\rangle
\label{cs}
\end{equation}
where ${\mathcal{N}}(\alpha)=
\exp(-\frac{1}{2}|\alpha|^2)$
is the normalization, and has all the well-known properties, such as minimizing the Heisenberg Uncertainty Principle.
\item {\bf Squeezed Vacuum} This state corresponds to the canonical transformation $U=S(\xi)$,
\begin{eqnarray}
a\rightarrow A = S(\xi) a S^{\dagger}(\xi) &=&\lambda a + \mu a^{\dagger} \nonumber \\
a^{\dagger}
\rightarrow
A^{\dagger} = S(\xi) a^{\dagger} S^{\dagger}(\xi) &=&\lambda^{*} a^{\dagger}+
\mu^{*} a
\end{eqnarray}
where $\lambda(\xi)$ and $\mu(\xi)$ satisfy $|\lambda |^2 - |\mu |^2 =1.$ Here $S(\xi)$ is the squeezing operator, given by
\begin{equation}
S (\xi) =\exp
\frac{1}{2}(
\xi ^{*}
a^2 - \xi a^{\dagger2}) ,  \;
\; \;
 (\xi = r\exp ( i\phi ))
\end{equation}
 where we have put
\[
\lambda= \cosh r, \; \; \;  \mu= \exp(i \phi ) \sinh r.
 \]
The squeezed vacuum satisfying $A|0'\rangle=0$ is given by $|\xi
\rangle  = S(\xi)|0\rangle$; that is, it satisfies
\begin{equation}
(\lambda a + \mu a^{\dagger})|\xi
\rangle =
0.
\end{equation}

The 2-parameter operator $S(\xi)$ is not the most general element of the 3-parameter vacuum-squeezing group $SU(1,1).$ An additional operator $P(\theta)=\exp(i\theta {{\hat{n}}})$, where ${{\hat{n}}}\equiv a^{\dagger}a ,$ supplies the third parameter and completes the generators of the group.
\item{\bf Squeezed States} Standard squeezed states correspond to the canonical transformation
 $U(\xi,\alpha)=S(\xi)D(\alpha)$ where
\begin{eqnarray}
a\rightarrow U(\xi,\alpha)a U^{\dagger}(\xi,\alpha)&=&\lambda a + \mu a^{\dagger} -\alpha I \nonumber \\
a^{\dagger}
\rightarrow
U(\xi,\alpha)a^{\dagger} U^{\dagger}(\xi,\alpha) &=&\lambda^{*} a^{\dagger}+
\mu^{*}
a-\alpha^{*} I \; \; \;
\; \; \;
|\lambda |^2 - |\mu |^2 =1.
\label{sg}
\end{eqnarray}
The squeezed state $ |\xi,\alpha\rangle = U(\xi,\alpha)|0\rangle$ thus defined may be written explicitly as
\begin{eqnarray}
|\xi,\alpha\rangle &=& U(\xi,\alpha)|0\rangle \nonumber \\
             &=& S(\xi)D(\alpha)|0\rangle \nonumber \\
             &=&\exp\frac{1}{2}( \xi ^{*} a^2 - \xi a^{\dagger 2})  \exp(\alpha a^{\dagger}-\alpha ^{*} a)|0\rangle
\end{eqnarray}
and satisfies the usual equation\cite{yuen}
\begin{equation}
(\lambda a + \mu a^{\dagger})|\xi,\alpha\rangle =\alpha |\xi,\alpha\rangle.
\end{equation}
 \end{itemize}
 We tabulate the above results in Table 1.  
 \begin{table}
  \begin{tabular}{|c|c|c|c|}
   \hline
   ${ A}$ & Unitary Operator ${ U}$ & EigenKet & State \\    \hline
   $a$ & $I$ (Identity) & $|0>$ & vacuum \\     \hline
   $a -\alpha I$& $D(\alpha)$  (Displacement Operator) & $|\alpha>$ & coherent state \\     \hline
   $\lambda a+ \mu a^{+}$ & $S(\lambda,\mu) $ (Squeezing Operator) & $|\xi>$ & squeezed vacuum \\     \hline
  $ \lambda a+ \mu a^{+}- \alpha I$ & $U(\lambda, \mu,\alpha)=S(\lambda,\mu)D(\alpha) $& $ |\xi,\alpha>$ &squeezed state  \\ \hline
 \end{tabular}
\caption{Canonical transformations and corresponding states}
\end{table}

The motivation for the foregoing observations is the following.  Since we have an explicit unitary transformation $U$ from the vacuum to a squeezed state, we have analogously the transformation from a free thermal state to our thermal squeezed state.  This enables any calculations involving the thermal squeezed state to be performed very easily.  And this may be expedited further by using the following  3-dimensional representation of the automorphism group of the Canonical Commutation relations generated by $U(\xi,\alpha)$, which we may call the {\em squeezing group} ${\mathcal{G}}$. This  is a 5-parameter group corresponding to the action of the semi-direct sum of the 3-dimensional Heisenberg-Weyl algebra (the central element of this algebra acts trivially) and the 3-dimensional algebra $su(1,1)$ on $\{a,a^{\dagger}\},$ and is a  subgroup of the inhomogeneous pseudounitary group $ISU(1,1)$ with fundamental representation
\begin{equation}
{\mathcal{G}}
\ni g =
\left[
\begin{array}{ccc}
\lambda&\mu&-\alpha\\
   \mu^{*}&\lambda^{*}&-\alpha^{*}  \\
0& 0&1
\end{array} \right]\; \; \; \; \; (\lambda,\mu,\alpha \in {\mathcal C}; \; \; |\lambda|^2-|\mu|^2=1).
 \end{equation}
 The value of this representation is due to the fact that for any operator $W$ the inverse transformed operator $U^{+}WU$ may be simply obtained by use of the inverse matrix $g^{-1}$
\begin{equation}\label{inverse}
g^{-1}=
\left[
\begin{array}{ccc}
\Lambda&M&-A\\
   M^{*}&\Lambda^{*}&-A^{*}  \\
0& 0&1
\end{array} \right]
\end{equation}
where
$$\Lambda=\lambda^{*}\; \; \; M=-\mu\; \; \; A=\mu \alpha^{*}-\lambda^{*}\alpha.$$
We shall make extensive use of this inverse transformation in the calculations  which follow.

 \section{Hamiltonian and Thermal Squeezed States}
 We define a thermal squeezed state quite conventionally as that obtained in the usual way from a hamiltonian.
 Our starting point is the single-mode  free boson hamiltonian (harmonic oscillator)
\begin{equation}\label{freeham}
 H_0=\epsilon (a^{\dagger}a+1/2); \; \; \; [a,a^{\dagger}]=1
\end{equation}
 with complete eigenspectrum $\{|n\rangle, n=0\ldots\infty\}$.
 We shall use the unitary transformation defined in the preceding section to produce the following
  hamiltonian with which to
 define our thermal states, writing the $\lambda, \mu$ dependence explicitly.
\begin{equation}\label{ham}
  H=U H_0 U^{\dagger}\; \; \; U=U(\lambda,\mu, \alpha)=S(\lambda,\mu)D(\alpha)
\end{equation}
where
\begin{equation}\label{trans}
U(\lambda,\mu,
\alpha) a
U^{\dagger}(\lambda,\mu,
\alpha)=\lambda
a+ \mu a^{\dagger}-\alpha.
\end{equation}
Explicitly
\begin{equation}\label{ham1}
H=\epsilon (\{\lambda a+ \mu a^{\dagger}-\alpha
\}^{\dagger}\{\lambda a+ \mu a^{\dagger}-\alpha \}+1/2).
\end{equation}
The eigenstates of $H$ are $U|n\rangle$ which are squeezed (displaced) number states. 

The Partition Function $Z$ for the Hamiltonian Eq.(\ref{ham1}) is just the standard
\begin{eqnarray}\label{pf}
Z = {\rm tr} e^{-\beta H} 
   = {\rm tr} e^{-\beta H_0} 
   =  e^{\frac{1}{2} x }(1-e^{x})^{-1} \; \; \; \; \; \; (\beta \equiv 1/kT \; \; \; x\equiv -\beta \epsilon)
\end{eqnarray} since the trace is invariant under the unitary transformation Eq.(\ref{trans}).
We define the thermal squeezed state $\rho_{TSS}$ by means of the Hamiltonian Eq.(\ref{ham1}) in the usual way
\begin{equation} \label{rho}
\rho_{TSS}=\exp(-\beta H)/Z.
\end{equation}
Thermal averages in the state $\rho_{TSS}$ are immediate due to the unitary equivalence with the free thermal state $\rho\equiv\exp(-\beta H_0)/Z$ defined in terms of the
 free boson Hamiltonian.
Generally, for the operator $W$, the thermal average is $$<W>_{TSS} = tr (\rho_{TSS}W)=<U^{+}WU>_{0}$$ where $<>_0$ is the free thermal average. As mentioned above, such thermal averages are simply evaluated using the inverse transformation $g^{-1}$ of Eq.(\ref{inverse}).
For example, the thermal averages of the quadratures are given by 
\begin{eqnarray}
 (\Delta X )^2 _{TSS} &=&|\lambda-\mu |^2 \;(\overline{n}+\frac{1}{2}
)  \label{avs1}\\
 (\Delta P )^2 _{TSS} &=&|\lambda+\mu |^2 \;(\overline{n}+\frac{1}{2} ) \label{avs2}
\end{eqnarray}
where $(\overline{n})$ is the average number of photons in the {\em free thermal state}. We obtain here the same  expressions as for the $n$-added squeezed state\cite{fs} with $\overline{n}$ substituted for $n$.
\section{Degradation of the Yuen Limit}
\begin{figure}
\vspace{1cm}
\begin{center}\resizebox{8 cm}{!}{\includegraphics{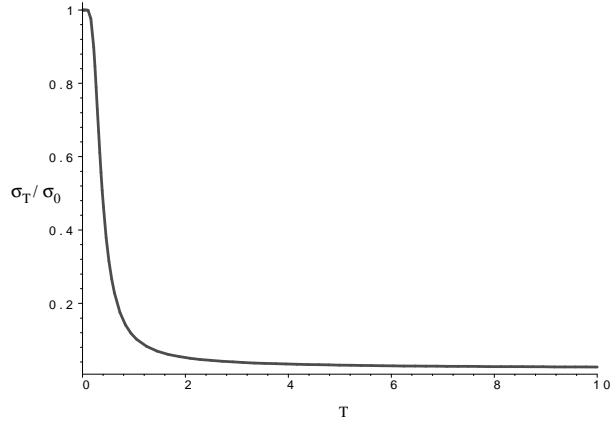}}
\caption{Signal-to-noise ratio {\it v}. Temperature}
\end{center}
\end{figure}
In an important yet elementary paper, Yuen\cite{yuen2} showed that the absolute optimal value for the signal-to-noise ratio $\sigma=<X>^2/(\Delta X)^2$ in a quantum system is given by $\sigma=4N(N+1)$, where $N$ is the photon intensity\footnote{This value is modified in non-standard quantum systems\cite{sol}.} (number of photons per unit time).   This limiting ratio is in fact attained by a standard squeezed state. 

It requires only  a straightforward calculation to evaluate the corresponding optimal ratio $\sigma_{T}$ in a thermal squeezed state, following the methods of \cite{fs} and \cite{yuen2} .  We assume that the maximum photon intensity $N$ is used, that all the signal is in one quadrature $X$, so that 
\begin{equation}\label{sigma}
\sigma=\{2N+1-(\Delta X)^2-(\Delta P)^2\}/(\Delta X)^2.
\end{equation}
We then optimize $\sigma$ with respect to $(\Delta X)^2$ subject to the constraint given by Eqs(\ref{avs1}, \ref{avs2})
\begin{equation}\label{sig2}
  (\Delta X)^2(\Delta P)^2=|\lambda^2-\mu^2|^2\;(\overline{n}+\frac{1}{2})^2.
\end{equation}
Eq.(\ref{sig2}) is minimal for real $\lambda,\mu$ , so that $|\lambda^2-\mu^2|^2=1$, giving
\begin{equation}\label{sigma2}
\sigma=(2N+1)/(\Delta X)^2-1-(\overline{n}+\frac{1}{2})^2/(\Delta X)^4.
\end{equation}
The optimal value  is given by
$$\sigma_T=\frac{4(N-\overline{n})(N+\overline{n}+1)}{(2\overline{n}+1)^2}$$
which value is attained for
\begin{eqnarray} 
\lambda&=&\frac{N+\overline{n}+1}{\sqrt{(2N+1)(2\overline{n}+1)}}\\
\mu&=&\frac{N-\overline{n}}{\sqrt{(2N+1)(2\overline{n}+1)}}\\
\alpha&=&\frac{(N+\overline{n}+1)(N-\overline{n})}{\sqrt{2\overline{n}+1}}
\end{eqnarray}

As noted,  $\overline{n}$ is the average number of photons in the {\em free thermal state}
 $$\overline{n}=e^{-\beta \epsilon}(1-e^{-\beta \epsilon})^{-1}.$$

In the Figure  we show how the Yuen limit is degraded by increasing temperature $T$ for our thermal squeezed states. ( We have normalized $\sigma_T$ to the optimal $T=0$ value $\sigma_0$, and measure $T$ in units of $\hbar \epsilon/k_B$   .) 
\section{Kraus model of thermalization}
Another advantage of the unitary transformation approach described above is that it enables us to relate the thermalization of squeezed states to that of the free boson vacuum state. Since we are not interested here in details of the dissipative evolution of our system, we eschew the more usual Lindblad differential equation approach and use  the global, Kraus formulation\cite{kraus}.  This gives the dissipative transformation of the state $\rho_0$ as
\begin{equation}\label{Kraus}
  \rho_0\longrightarrow \rho= \sum_n {w_n \rho_0 w_{n}^{\dagger}}
\end{equation}
where $\sum_n{w_n^{\dagger}w_n}=I$ and the operators (matrices) $w_n$ are otherwise arbitrary.
It is clear that this preserves both the positivity of  $\rho_0$ and the  tr $\rho_o=1$ property.

(Perhaps less obvious is the fact that the set $S$ of transformation sets $\{w_n\}$ gives rise to a semi-group, since for any two elements  $g=\{w_n\}$ and $g'=\{w'_n\}$  the set product $gg'$ also satisfies the Kraus conditions, $gg' \in S$. The semi-group $S$ has an identity, but the only elements with inverses are  the single-element sets $\{U\}$, where $U$ is unitary.)  

As a simple 2-level illustration of the Kraus procedure, we may describe the dissipation of the pure state
 $\rho_0=
 \left[
\begin{array}{cc}
 1 & 0 \\
 0 & 0 
\end{array}
\right]$
to the mixed state
 $\rho=
  \left[
\begin{array}{cc}
 p_0 & 0 \\
 0   & p_1 
\end{array}
\right]$
by
\begin{equation}\label{Kraus2}
  \rho_0\longrightarrow \rho= \sum_{n=0}^1{w_n \rho_0 w_{n}^{\dagger}}
\end{equation}
with 
 $$
 w_0=\sqrt{p}
 \left[
\begin{array}{cc}
 1 & 1-p \\
 0 & \sqrt{2p-p^2} 
\end{array}
\right]
\; \; \; 
w_1=\sqrt{1-p}
 \left[
\begin{array}{cc}
 0 & \sqrt{1-p^2} \\
 1 & -p 
\end{array}
\right]
$$
Of course the Kraus transformation set $\{ w_n \}$ is not unique.

Perhaps the general pattern of this Kraus set will emerge more easily if we write down the $3\times 3$ case.  The set of matrices $\{w_n\}$ which transform 
$$\rho_0 = \left[\begin{array}{ccc}1&0&0\\0&0&0\\0&0&0
\end{array}
\right]
\longrightarrow
\rho = \left[\begin{array}{ccc}p_0&0&0\\0&p_1&0\\0&0&p_2
\end{array}
\right]$$
is given by:
$$
w_0=
\sqrt{p_0}\left[
\begin{array}{ccc}
 \begin{array}{c}1\\0\\0 \end{array} \left|\begin{array}{c} \mbox{}  \underline{p^0_1} \\ \mbox{} \end{array}\right| \begin{array}{c}  \mbox{}  \underline{p^0_2} \\\mbox{}  \end{array}
\end{array}
\right]
\; \; 
w_1=
\sqrt{p_1}\left[
\begin{array}{ccc}
 \begin{array}{c}0\\1\\0 \end{array} \left|\begin{array}{c} \mbox{}  \underline{p^1_1} \\ \mbox{} \end{array}\right| \begin{array}{c}  \mbox{}  \underline{p^1_2} \\\mbox{}  \end{array}
\end{array}
\right]
\; \;
w_2=
\sqrt{p_2}\left[
\begin{array}{ccc}
 \begin{array}{c}0\\0\\1 \end{array} \left|\begin{array}{c} \mbox{}  \underline{p^2_1} \\ \mbox{} \end{array}\right| \begin{array}{c}  \mbox{}  \underline{p^2_2} \\\mbox{}  \end{array}
\end{array}
\right]
$$
where the vectors $\underline{p^n_u}$ satisfy the orthonormality relations
$$\underline{p^n_r}\, . \,\underline{p^n_s}=\delta_{rs}\;\; \; \; \; (r,s >0),$$
 as well as the probability conditions $\sum_n{p_n p^n_{n,r}}=0\; \; \; (r>0).$
Note that we are numbering our matrix elements from $(0,0)$.

In the general case the conditions on $w_n=\sqrt{p_n}\sum_{r,s}p^n_{r,s} |r \rangle\langle s|\; \; \; (n,r,s=0,\ldots )$ are
\begin{enumerate}
\item Orthonormality: $\sum {{p^n_{r,s}}^{\dagger}\;p^n_{r,v}}=\delta_{s,v}\; \; \; n=0,1,\ldots, s,v >0$
\item Probability matching:  $p^n_{r,0}=\delta^n_r$
\item Probability orthogonality: $\sum_n{p_n p^n_{n,r}} = 0\; \; \; r>0$
\end{enumerate}

The foregoing permits an immediate description  of the thermalization process for a squeezed state.  Consider a Kraus set $\{w_n\}$ which thermalizes the pure state 
$\rho_0\equiv |0\rangle\langle0|$ giving the free thermal state $\rho=e^{-\beta H_0}/Z$
$$ \rho_0\stackrel{K}\longrightarrow \rho = \sum_n{w_n \rho_0 w_n^{\dagger}}.$$
Then the pure squeezed state $\rho_S =|\xi,\alpha\rangle\langle\xi,\alpha|$ dissipates to the  thermal squeezed state $\rho_{TSS}=e^{-\beta H}/Z=U \rho U^{\dagger}$ using the Kraus set $\{\tilde{w_n}\}$,
$$ \rho_{S}\stackrel{K}\longrightarrow \rho_{TSS} = \sum_n{\tilde{w_n} \rho_S \tilde{w_n}^{\dagger}}$$
where $\tilde{w_n}=U w_n U^{\dagger}$.
\section{Conclusion}
We have shown that using  a group theoretic definition of squeezed states, it is a straightforward matter to define a thermal squeezed state, which lends itself simply to evaluation of physical quantities.  We exemplified this by evaluating the Yuen limit for the signal-to-noise ratio in such a thermal squeezed state as a function of the temperature $T$, showing that it is reduced below the optimal $T=0$ value. Finally, we have outlined  a general model for the thermalization of the standard squeezed state using the Kraus formalism. 

\end{document}